\documentclass[acus]{JAC2003}
\usepackage{graphicx}
\begin{document}

\thispagestyle{empty}

\onecolumn

\begin{flushright}
{\large
SLAC--PUB--9812\\
May 2003\\}
\end{flushright}

\vspace{.8cm}

\begin{center}

{\LARGE\bf
Tracking Simulations Near Half-Integer Resonance at PEP-II~\footnote
{\normalsize
{Work supported by Department of Energy contract  DE--AC03--76SF00515.}}}

\vspace{1cm}

\large{
Y.~Cai and Y.~Nosochkov\\
Stanford Linear Accelerator Center, Stanford University,
Stanford, CA 94309}

\end{center}

\vfill

\begin{center}
{\LARGE\bf
Abstract }
\end{center}

\begin{quote}
\large{
Beam-beam simulations predict that PEP-II luminosity can be increased by
operating the horizontal betatron tune near and above a half-integer
resonance.  However, effects of the resonance and its synchrotron sidebands
significantly enhance betatron and chromatic perturbations which tend to
reduce dynamic aperture.  In the study, chromatic variation of horizontal
tune near the resonance was minimized by optimizing local sextupoles in the
Interaction Region.  Dynamic aperture was calculated using tracking
simulations in LEGO code.  Dependence of dynamic aperture on the residual
orbit, dispersion and $\beta$ distortion after correction was investigated.
}
\end{quote}

\vfill

\begin{center}
\large{
{\it Presented at the 2003 Particle Accelerator Conference
(PAC 03)\\
Portland, Oregon, May 12--16, 2003}
} \\
\end{center}

\newpage

\pagenumbering{arabic}
\pagestyle{plain}

\twocolumn

\title{
TRACKING SIMULATIONS NEAR\\
HALF-INTEGER RESONANCE AT PEP-II~\thanks
{Work supported by Department of Energy contract
DE--AC03--76SF00515.}\vspace{-4mm}}

\author{
Y.~Cai, Y.~Nosochkov,
SLAC, Menlo Park, CA 94025, USA}

\maketitle

\begin{abstract}

Beam-beam simulations predict that PEP-II luminosity can be increased by
operating the horizontal betatron tune near and above a half-integer
resonance.  However, effects of the resonance and its synchrotron sidebands
significantly enhance betatron and chromatic perturbations which tend to
reduce dynamic aperture.  In the study, chromatic variation of horizontal
tune near the resonance was minimized by optimizing local sextupoles in the
Interaction Region.  Dynamic aperture was calculated using tracking
simulations in LEGO code.  Dependence of dynamic aperture on the residual
orbit, dispersion and $\beta$ distortion after correction was investigated.

\end{abstract}

\section{INTRODUCTION}

\begin{figure}[tb]
\centering
\includegraphics*[width=82mm]{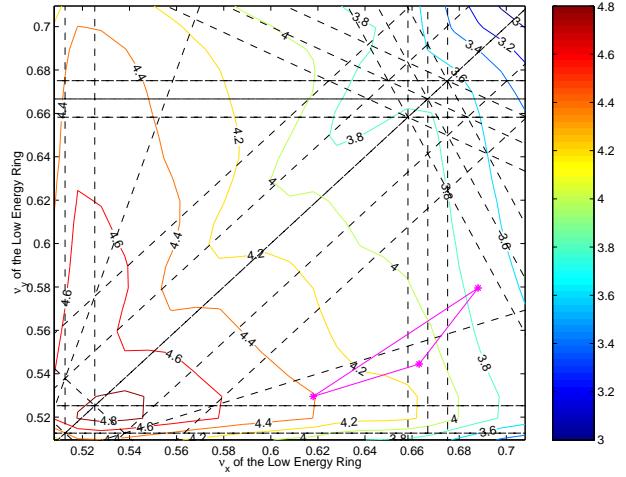}
\vspace{-6mm}
\caption{Single bunch luminosity scan [$10^{30}$~cm$^{-2}$s$^{-1}$].}
\label{fig:lumin}
\vspace{-0mm}
\end{figure}

PEP-II~\cite{cdr} has been operating at the betatron tune $\nu_x/\nu_y$
close to $24.569/23.639$ in the High Energy Ring (HER) and $38.649/36.564$
in the Low Energy Ring (LER).  These working points were selected
experimentally for a reliable machine performance, good luminosity and beam
lifetime.  However, the beam-beam simulations predict that luminosity can
be increased by operating betatron tune very close and above the
half-integer resonance.  Fig.~\ref{fig:lumin} shows the LER tune diagram
with synchro-betatron resonances up to the 4th order and a contour plot of
the single bunch luminosity.  Calculation of luminosity was done using the
beam-beam code developed at SLAC~\cite{bbeam} which has been recently
upgraded to the three dimensional version.

The difficulty of operating close to half-integer resonance comes from
enhancement of the resonance effects on betatron motion.  It is
well known that perturbation of $\beta$ function created by focusing errors
depends on tune $\nu$ as
\begin{equation}
\frac{\Delta\beta}{\beta}(s)=\frac{1}{2\sin2\pi\nu}
\oint \beta(l) \Delta K_1(l) \cos2\phi(s,l) dl,
\label{eqn:dbeta}
\end{equation}
where $\mu$ is phase advance,
$\phi(s,l)\!=\!\pi\nu\!-\!\left|\mu(s)\!-\!\mu(l)\right|$, and $\Delta K_1$
is a focusing error created mainly by quadrupole field imperfections,
horizontal orbit at sextupoles, and momentum error.  Close to half-integer
resonance, growth of $\Delta\beta/\beta$ comes from the resonance term
$[\sin2\pi\nu]^{-1}$ which behaves as $1/\Delta\nu$ when distance to the
resonance is as small as $\Delta\nu\!\ll\!1/2\pi$.  On the other hand,
orbit and dispersion are not excited by the half-integer resonance.

For significant enhancement of luminosity, fractional value of horizontal
tune should be in the range of $[\nu_x]\!\approx\!.51$.  At this working
point, enhancement of $\Delta\beta_x/\beta_x$ in HER and LER due to the
resonance term in Eqn.~\ref{eqn:dbeta} would be a factor of 6.7 and 12.8,
respectively, compared to the present tune.  Without compensation, the
large $\beta$ growth may significantly increase amplitude dependent
non-linear aberrations and reduce dynamic aperture and beam lifetime.

More resonance effects are generated by the synchrotron sidebands of
the half-integer resonance: $2\nu_x\!+\!m\nu_s\!=\!n$,
where $\nu_s$ is a synchrotron tune, and $m, n$ are integers.  In the LER,
where $\nu_s\!=\!0.025$, the 1st and 2nd synchro-betatron resonances occur
at $[\nu_x]\!=\!.5125$ and .525, while in HER with $\nu_s\!=\!0.045$ the
1st sideband is at $[\nu_x]\!=\!.5225$.  Tracking simulations will show
that the sidebands have a strong effect on dynamic aperture, therefore
working tune should be chosen reasonably far from them.  In
addition, variation of tune with synchrotron momentum oscillations should 
be minimized to avoid crossing with these resonances.

Optimization of PEP-II lattice near half-integer resonance and analysis of
dynamic aperture are discussed below.  The optics with
$\beta_x^*/\beta_y^*\!=\!50/1.25$~cm at the Interaction Point (IP) is used.

\section{LATTICE OPTIMIZATION}

PEP-II has two tuning sections which can be locally adjusted to change
betatron tune without affecting the rest of machine optics.  Initially,
only these sections were modified to move the horizontal tune closer to
half-integer, and vertical tune to $[\nu_y]\!\approx\!.61$ as suggested by
beam-beam analysis.  But tracking simulations showed that dynamic aperture
was not sufficiently large with machine errors and synchrotron momentum
oscillations of up to $\pm8\sigma_p$, where $\sigma_p$ is the {\it rms}
value of relative momentum spread $\frac{\Delta p}{p}$ in the beam.
Analysis of chromaticity indicated that non-linear variation of horizontal
tune with momentum needs to be further reduced to avoid
crossing with the synchro-betatron resonances.

In PEP-II, the most contribution to non-linear chromaticity is generated in
the final quadrupole doublets near IP.  This chromaticity is compensated by
the Interaction Region (IR) sextupoles located in the same phase with the
doublets.  Variation of strength of these sextupoles allows to compensate
quadratic dependence of tune on $\frac{\Delta p}{p}$, and a small adjustment
of sextupole phase advance helps reduce the higher order variation.

Minimum of the second order chromaticity was achieved by reducing strengths
of the IR sextupoles correcting horizontal chromaticity.  Further
improvement in LER resulted from reduction of horizontal phase advance
between the IR horizontal sextupoles and IP by $5^\circ$.  For correction
of the machine linear chromaticity, strength of the global sextupoles was
increased to compensate for the weaker IR sextupoles.  Because the adjusted
IR sextupoles in LER have a non-zero design orbit, the reduced sextupole
strength created a feed-down effect of linear focusing and coupling.  This
small perturbation was compensated by a slight adjustment of the IR magnet
strengths.

\begin{figure}[tb]
\centering
\includegraphics*[width=82mm]{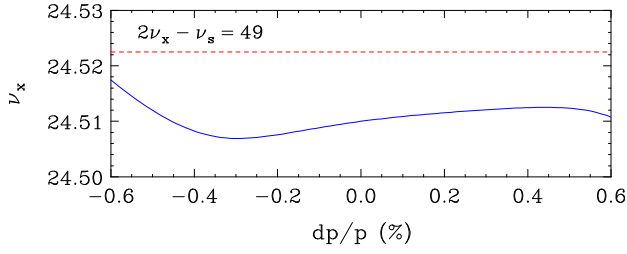}
\vspace{-6mm}
\caption{HER horizontal tune vs. $\frac{\Delta p}{p}$ at $\nu_x\!=\!24.51$.}
\label{fig:htunex}
\vspace{-0mm}
\end{figure}

\begin{figure}[tb]
\centering
\includegraphics*[width=82mm]{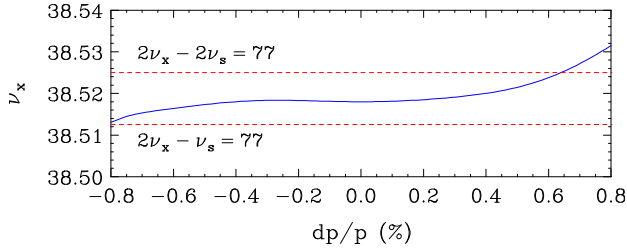}
\vspace{-6mm}
\caption{LER horizontal tune vs. $\frac{\Delta p}{p}$ at
$\nu_x\!=\!38.518$.}
\label{fig:ltunex}
\vspace{-4mm}
\end{figure}

The optimized horizontal tune for momentum range of
$-10\sigma_p\!<\!\frac{\Delta p}{p}\!<\!10\sigma_p$ is shown in
Fig.~\ref{fig:htunex},~\ref{fig:ltunex}, where the working point is
$\nu_x/\nu_y\!=\!24.51/23.61$ in HER and $38.518/36.61$ in LER.  The
straight dash lines depict the half-integer synchrotron sidebands.  In HER,
a positive linear chromaticity $\xi\!=\!+1$ was used in
Fig.~\ref{fig:htunex} to counteract the negative slope of non-linear tune
variation.  Due to the large synchrotron tune, it was possible to place the
HER working point below the 1st synchrotron sideband without crossing with
the resonance lines.  In the LER, synchrotron tune is a factor of 2 smaller
while energy spread is 25\% larger, therefore the closest to half-integer
working point was chosen between the 1st and 2nd sidebands.

\section{DYNAMIC APERTURE SIMULATIONS}

\begin{figure}[tb]
\centering
\includegraphics*[width=82mm]{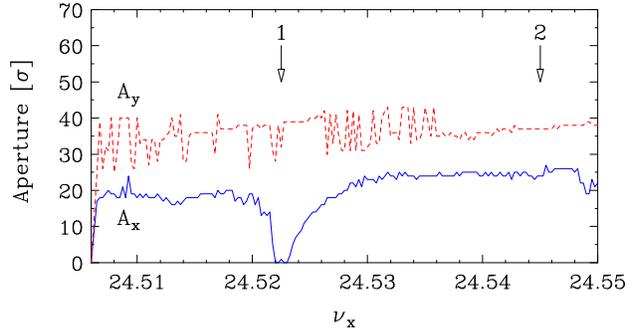}
\vspace{-7mm}
\caption{HER dynamic aperture without magnet errors vs. $\nu_x$
at $\nu_y\!=\!23.61$. Synchrotron sidebands:
1) $2\nu_x\!-\!\nu_s\!=\!49$, 2) $2\nu_x\!-\!2\nu_s\!=\!49$.}
\label{fig:hscanx}
\vspace{-0mm}
\end{figure}

\begin{figure}[tb]
\centering
\includegraphics*[width=82mm]{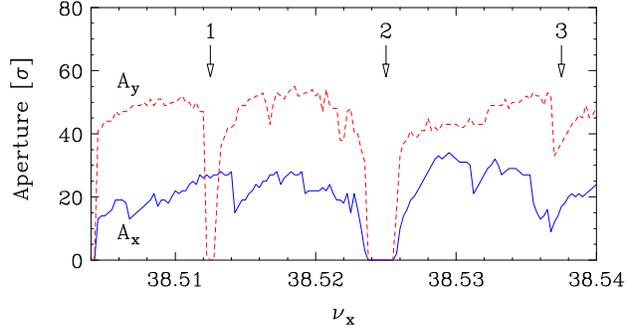}
\vspace{-7mm}
\caption{LER dynamic aperture without magnet errors vs. $\nu_x$
at $\nu_y\!=\!36.61$. Synchrotron sidebands:
1) $2\nu_x\!-\!\nu_s\!=\!77$, 2) $2\nu_x\!-\!2\nu_s\!=\!77$,
3) $2\nu_x\!-\!3\nu_s\!=\!77$.}
\label{fig:lscanx}
\vspace{-3mm}
\end{figure}

Calculations of dynamic aperture were performed using tracking simulations
in LEGO code~\cite{lego}.  First, dependence of aperture on betatron tune
near half-integer resonance was investigated for lattice without magnet
errors, but with synchrotron momentum oscillations of $\pm8\sigma_p$.  The
resultant horizontal tune scan is shown in Fig.~\ref{fig:hscanx},
\ref{fig:lscanx} for HER and LER, where dynamic aperture is normalized by
the {\it rms} size of a fully coupled beam.

The HER horizontal dynamic aperture vanishes in the vicinity of the main
half-integer resonance $2\nu_x\!=\!49$ and its 1st sideband.  In LER, the
strongly affecting resonances are $2\nu_x\!+\!m\nu_s\!=\!77$ with
$m\!=\!0,-1,-2,-3$.  Naturally, the working tune should be chosen
reasonably far from these resonances.  The following scenarios for more and
less aggressive horizontal tune $\nu_x$ near half-integer were investigated
in the simulations:

\begin{enumerate}
\vspace{-3mm}
\item HER: 24.51, \hspace{4.7mm} LER: 38.518.
\vspace{-3mm}
\item HER: 24.529, \hspace{3mm} LER: 38.529.
\vspace{-3mm}
\end{enumerate}

Dynamic aperture scan versus vertical tune was performed for the first
scenario of $\nu_x$ and the range of $[\nu_y]$ from .55 to .64.  It showed
that dynamic aperture gradually reduces as $[\nu_y]$ becomes closer to
$[\nu_x]$ and the working point approaches the crossing of half-integer and
coupling resonances.  Based on this scan, the vertical fractional tune of
$[\nu_y]\!=\!.61$ was chosen for these simulations.  A lower $\nu_y$ may
be considered for further luminosity enhancement.

Secondly, tracking simulations with field errors, misalignment and
$\frac{\Delta p}{p}\!=\!\pm8\sigma_p$ synchrotron momentum oscillations
were performed for the selected working points.  For statistics, ten
different settings (``seeds'') of random machine errors were used in each
tracking.  Perturbation of beam orbit, linear chromaticity, betatron tune
and vertical dispersion was compensated using realistic correction schemes
in LEGO.  Since distortion of $\beta$ function becomes more sensitive to
focusing errors near half-integer resonance, a special correction of
$\frac{\Delta \beta}{\beta}$ was implemented in LEGO.  It uses MICADO
method to find the most effective quadrupoles to minimize $\beta$
perturbation.

Due to the greater effect of errors near half-integer resonance, a better
machine correction is needed to maintain acceptable dynamic aperture.  To
verify tolerance to various errors, simulations were performed for
different levels of machine correction.  It has been confirmed that beam
orbit should be decreased for an acceptable dynamic aperture.  The better
orbit correction reduces the feed-down focusing errors in sextupoles as
well as residual dispersion in the machine.  On the other hand, correction
of vertical dispersion did not significantly affect dynamic aperture in the
observed range of residual {\it rms} $\Delta\eta_y$ from $\sim$70 to 5~mm.

As expected, the simulations confirmed that compensation of $\Delta
\beta/\beta$ is required in the first scenario, where $\nu_x$ is closer to
the resonance.  Typically, $\Delta \beta/\beta$ was corrected to the {\it
rms} level of $<\!5\%$.  In the second scenario, at $[\nu_x]\!=\!.529$,
correction of $\beta$ function was less important, although it helped to
improve cases with small aperture.  Table~1 summarizes the observed
approximate levels of {\it rms} orbit and $\Delta\beta_x/\beta_x$ for
acceptable dynamic aperture.

\begin{table}[htb]
\small
\begin{center}
\vspace{-3mm}
\caption{Tolerances on {\it rms} orbit and $\Delta\beta_x/\beta_x$.}
\medskip
\begin{tabular}{|c|c|c|c|c|}
\hline
 & \multicolumn{2}{c|}{\textbf{HER}} & \multicolumn{2}{c|}{\textbf{LER}} \\
\hline
\boldmath{$[\nu_x]$} & \textbf{.529} & \textbf{.510} &
\textbf{.529} & \textbf{.518} \\
\hline
orbit (mm) & 1 & 1 & 0.4 & 0.2 \\
$\Delta \beta_x/\beta_x$ (\%) & 25 & 5 & 25 & 5 \\
\hline
\end{tabular}
\label{tab:toler}
\end{center}
\vspace{-3mm}
\end{table}

\begin{figure}[tb]
\centering
\includegraphics*[width=60mm]{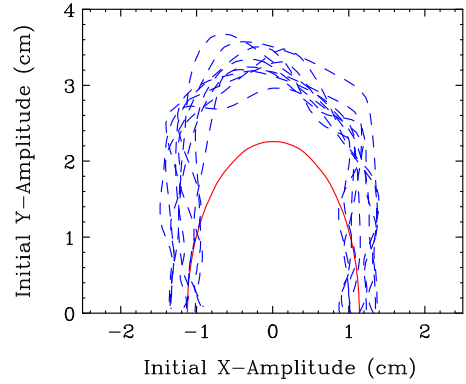}
\vspace{-3mm}
\caption{HER dynamic aperture at
$\nu_x/\nu_y\!=\!24.51/23.61$.}
\label{fig:haper51}
\vspace{1mm}
\end{figure}

The resultant dynamic aperture in HER at $\nu_x/\nu_y\!=\!24.51/23.61$ for
a good level of correction is shown in Fig.~\ref{fig:haper51} at the
injection point.  Stable particle motion corresponds to the area inside the
dash lines which represent 10 different seeds of random machine errors.
The solid half-ellipse, shown for reference, is the $10\sigma$ size of a
fully coupled beam at injection with emittance $\epsilon_x\!=\!48$~nm and
$\epsilon_y\!=\!\epsilon_x/2$.  In this simulation, the residual {\it rms}
orbit, dispersion and $\beta$ distortions after correction were:
$0.27/0.31$~mm, $\Delta\eta\!=\!31/8$~mm and
$\Delta\beta/\beta\!=\!1.9/1.5$\% in $x/y$ planes, respectively.  Linear
chromaticity was set to +1 to minimize non-linear tune variation with
momentum.

\begin{figure}[tb]
\centering
\includegraphics*[width=60mm]{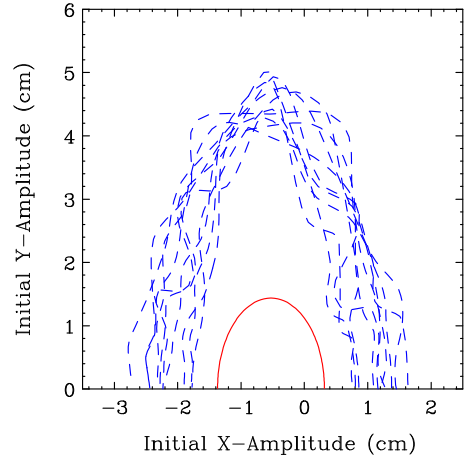}
\vspace{-3mm}
\caption{LER dynamic aperture at
$\nu_x/\nu_y\!=\!38.518/36.61$.}
\label{fig:laper518}
\vspace{-3mm}
\end{figure}

The LER dynamic aperture at $\nu_x/\nu_y\!=\!38.518/36.61$ is shown in
Fig.~\ref{fig:laper518} at the injection point.  The solid half-ellipse
corresponds to $10\sigma$ size of a fully coupled beam at injection with
emittance $\epsilon_x\!=\!24$~nm.  The residual {\it rms} orbit, dispersion
and $\beta$ distortions after correction were:  $0.20/0.23$~mm,
$\Delta\eta\!=\!23/9$~mm and $\Delta\beta/\beta\!=\!2.4/2.5$\% in $x/y$
planes, respectively.  Linear chromaticity was set to zero in this case.

Implementation of the tune near half-integer resonance has been recently
performed at PEP-II.  The working point was successfully moved to
$\nu_x/\nu_y\!=\!24.52/23.63$ in HER and $38.52/36.57$ in LER.  After the
necessary machine adjustments luminosity has been improved by $\sim$15\% to
the new record of $6.1\!\cdot\!10^{33}$~cm$^{-2}$s$^{-1}$.

\vspace{-1mm}
\section{CONCLUSION}

Beam-beam simulations performed for PEP-II predicted an enhancement of
luminosity for a betatron tune near a half-integer resonance.  Horizontal
fractional tune of .52 has been recently implemented at PEP-II and
$\sim$15\% luminosity gain has been achieved.  Particle tracking
simulations showed that an improved machine correction is needed for
acceptable dynamic aperture at $[\nu_x]\!=\!.51$ in HER and .518 in LER.
This requires a minimization of non-linear chromaticity and tighter
correction of orbit and $\beta_x$ distortions.  At $[\nu_x]\!=\!.529$, a
looser orbit correction may be used while compensation of
$\Delta\beta_x$ may not be necessary.


\end{document}